\documentclass[aps,prl,twocolumn,superscriptaddress,showpacs]{revtex4-1}
\usepackage{graphicx}
\usepackage{mathrsfs}
\usepackage{bm}
\usepackage{amsmath}
\usepackage{dcolumn}
\usepackage{epstopdf}
\usepackage{dsfont}
\usepackage{amssymb}
\usepackage{tabularx}
\usepackage{array}
\usepackage{braket}
\usepackage{xcolor,colordvi}
\usepackage{txfonts}

\begin{document}
\title{Spin-Pairing Correlations and Spin Polarization of Majorana Bound States in Two-Dimensional Topological Insulator Systems}
\author{Kunhua Zhang}
\affiliation{ICQD, Hefei National Laboratory for Physical Sciences at Microscale, and Synergetic Innovation Centre of Quantum Information and Quantum Physics, University of Science and Technology of China, Hefei, Anhui 230026, China}
\affiliation{CAS Key Laboratory of Strongly-Coupled Quantum Matter Physics and Department of Physics, University of Science and Technology of China, Hefei, Anhui 230026, China}
\author{Junjie Zeng}
\affiliation{ICQD, Hefei National Laboratory for Physical Sciences at Microscale, and Synergetic Innovation Centre of Quantum Information and Quantum Physics, University of Science and Technology of China, Hefei, Anhui 230026, China}
\affiliation{CAS Key Laboratory of Strongly-Coupled Quantum Matter Physics and Department of Physics, University of Science and Technology of China, Hefei, Anhui 230026, China}
\author{Yafei Ren}
	\affiliation{ICQD, Hefei National Laboratory for Physical Sciences at Microscale, and Synergetic Innovation Centre of Quantum Information and Quantum Physics, University of Science and Technology of China, Hefei, Anhui 230026, China}
	\affiliation{CAS Key Laboratory of Strongly-Coupled Quantum Matter Physics and Department of Physics, University of Science and Technology of China, Hefei, Anhui 230026, China}
\author{Zhenhua Qiao}
\email[Correspondence author:~]{qiao@ustc.edu.cn}
\affiliation{ICQD, Hefei National Laboratory for Physical Sciences at Microscale, and Synergetic Innovation Centre of Quantum Information and Quantum Physics, University of Science and Technology of China, Hefei, Anhui 230026, China}
\affiliation{CAS Key Laboratory of Strongly-Coupled Quantum Matter Physics and Department of Physics, University of Science and Technology of China, Hefei, Anhui 230026, China}

\begin{abstract}

We demonstrate that zero-energy Majorana bound state, in the ferromagnetic insulator (FI)-superconductor (SC) junction formed on the edge of two-dimensional topological insulator, exhibits three types of spin-triplet pairing correlations and its spin polarization direction is position independent in ferromagnetic insulator. When an electron is injected with a spin (anti-)parallel to this direction, equal-spin Andreev reflection exhibits the widest (narrowest) resonance peak. Similar behaviour is found when the coupling between two Majorana bound states in a FI-SC-FI junction is invoked, though an additional weak spin-singlet pairing correlation is generated. These signatures can readily facilitate the experimental detection of spin-triplet correlations and spin polarization of Majorana bound states.
\end{abstract}

\maketitle

\emph{Introduction---.} Majorana fermions are exotic particles that are their own antiparticles~\cite{F.Wilczek2009}, and have been suggested to exist as Majorana bound states (MBSs) in condensed matter systems~\cite{S.Elliott2015}. Two spatially separated MBSs can define a qubit that stores information non-locally and is robust against local sources of decoherence~\cite{A.Kitaev2001}, which together with its non-Abelian statistics~\cite{D.Ivanov2001,C.Nayak2008} make it exhibit potential applications in quantum information and quantum computation~\cite{A.Kitaev2003}. Several theoretical proposals were raised to realize such states, like topological insulators proximity-coupled with superconductors~\cite{L.Fu2008, L.Fu2009a, X.Qi20010, Beenakker2013a}, semiconductor-superconductor heterostructures~\cite{J.Sau2010,J.Alicea2010,R.Lutchyn2010,Y.Oreg2010}, and magnetic-atomic chains on superconductors~\cite{N.Perge2012}. Recently, intensive theoretical and experimental efforts have been made to verify the existence of MBSs in these systems by employing charge transport properties~\cite{C.Bolech2007, J.Nilsson2008, L.Fu2009b, A.R.Akhererov2009, K.T.Law2009, K.Flensberg2010, D.Pikulin2012, E.Prada2012, E.Lee2012, V.Mourik2012, M.Deng2012,J.Liu2012,B.H.Wu2012}.
However, few attention has been paid to the spin related phenomena of MBSs~\cite{H.Ebisu2015, J.J.He2014a, A.Haim2015, H.H.Sun2016}. Furthermore, the in-depth classification of spin-triplet correlations and spin polarization of MBSs are yet unclear~\cite{X.Liu2015, N.Yuan2015a}, especially those in two-dimensional topological insulator systems. And these characteristics are closely related to the resulting unusual spin-related transport, like the intriguing selective equal-spin Andreev reflection~\cite{J.J.He2014a, H.H.Sun2016}.

In this Letter, we present a systematic study of spin-pairing correlations and spin polarization of MBS/MBSs in ferromagnetic insulator(FI)-superconductor(SC) and FI-SC-FI junction formed at the boundary of a two-dimensional topological insulator. For the FI-SC junction, we find a zero-energy MBS, which possesses three types of spin-triplet pairing correlations and its spin polarization orientation remains unchanged in the ferromagnetic insulator regime. When two MBSs are coupled in the FI-SC-FI junction, an additional weak spin-singlet pairing correlation is generated. In both cases, the dominated spin-triplet correlations induce strongly contrasted widths of equal-spin Andreev reflection peaks for injected electrons with different spin polarizations. As a consequence, the spin-pairing correlations and spin polarization of MBSs could be experimentally detected in the spin-related transport measurements.

\begin{figure}
\includegraphics[width=8cm]{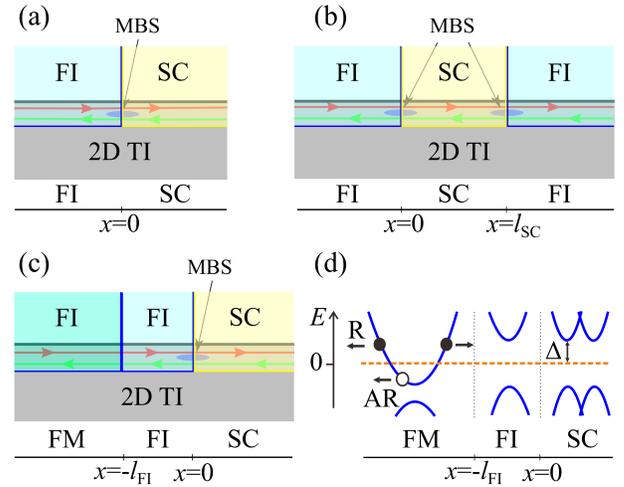}
\caption{(color online) (a-c) Schematics of one-dimensional FI-SC, FI-SC-FI and ferromagnetic metal (FM)-FI-SC junctions mediated by the edge states of two-dimensional topological insulator (2D TI) systems, respectively. (d) Schematic energy band for junction in (c). Solid and open circles indicate electrons and holes, respectively. R and AR indicate the electron reflection and Andreev reflection, respectively.}
\label{fig1}
\end{figure}

\emph{Model and Formalism---.} We consider two different one-dimensional setups, which are FI-SC and FI-SC-FI junctions formed at the boundary of a two-dimensional topological insulator as displayed in Figs.~\ref{fig1}(a) and \ref{fig1}(b). The one-dimensional edge states proximity-coupled with bulk ferromagnetic insulator and $s$-wave superconductor~\cite{L.Fu2008} can be described by the following Bogoliubov-de Gennes equation in the representation spanned on the basis of $\{\varphi_{\uparrow}, \varphi_{\downarrow}, \varphi^{\dag}_{\downarrow}, -\varphi^{\dag}_{\uparrow}\}$~\cite{L.Fu2009a, Beenakker2013a}:
\begin{align}
	\label{BdG Hamiltonian}
	\begin{pmatrix}
		\upsilon _{\rm {F}} \sigma_{x} p_{x}+{\boldsymbol{\sigma}}\cdot{\boldsymbol{m}}-\mu&\Delta \mathrm{e}^{\mathrm{i}\phi}\\
		\Delta \mathrm{e}^{-\mathrm{i}\phi}&-\upsilon_\text{F}\sigma_{x} p_{x}+\boldsymbol{\sigma}\cdot \boldsymbol{m}+\mu
	\end{pmatrix} \psi = E\psi,
\end{align}
where $\boldsymbol{\sigma}=(\sigma_{x},\sigma_{y},\sigma_{z})$ and $\upsilon_{\rm F}$ are respectively Pauli matrices and Fermi velocity of the topological-insulator edge states. The proximity effects are reflected by the magnetization $\boldsymbol{m}$ and pair potential $\Delta \mathrm{e}^{\mathrm{i}\phi}$ that occur only at the ferromagnetic insulator and superconductor regimes separately. In Fig.~\ref{fig1}(a), the magnetization $\boldsymbol{m}$ is set to be $(0,0,m_\text{L})$, and in Fig.~\ref{fig1}(b) it is set to be $(0,0,m_\text{L/R})$ at the left/right sides of the superconductor. In our calculations, the phase $\phi$ of pair potential plays no role and thus is set to be zero in below. The chemical potential $\mu(x)$ is determined with respect to the Dirac point and is assumed to be independently tunable via gating or doping in each regime~\cite{I.Knez2012}.

By solving Eq.~(\ref{BdG Hamiltonian}), one can obtain the wavefunctions in the junctions shown in Fig.~\ref{fig1}, e.g., the wavefunction in the left ferromagnetic insulator region of Fig.~\ref{fig1}(a) is $\psi_\text{FI}(x)=a_\text{e}(-\hbar\upsilon_\text{F}k^{+}_\text{FI},E+\mu_\text{FI}-m_\text{L},0,0)^{T}\mathrm{e}^{-\mathrm{i}k^{+}_\text{FI}x}
+a_\text{h}(0,0,\hbar\upsilon_\text{F}k^{-}_\text{FI},E-\mu_\text{FI}-m_\text{L})^{T}\mathrm{e}^{-\mathrm{i} k^{-}_\text{FI}x}$
where $ k^{\pm}_\text{FI}=\mathrm{i} \sqrt{m_\text{L}^2-(\mu_\text{FI}\pm E)^2}/\hbar\upsilon_\text{F} $, $ \mu_\text{FI} $ is the chemical potential, $ a_\text{e/h} $ are the coefficients of evanescent wavefunctions for electron and hole, respectively. And the wavefunction in the right superconducting region is
$\psi_\text{SC}(x)=b(-\mathrm{e}^{-\mathrm{i} \alpha},\mathrm{e}^{-\mathrm{i} \alpha},-1,1)\mathrm{e}^{-\mathrm{i}k_\text{SC}x-Kx}
+c(\mathrm{e}^{\mathrm{i} \alpha},\mathrm{e}^{\mathrm{i}\alpha},1,1)^{T}\mathrm{e}^{\mathrm{i} k_\text{SC}x-Kx}$
where $k_\text{SC}=\mu_\text{SC}/\hbar\upsilon_\text{F}$, $K={\Delta\sin\alpha}/{\hbar\upsilon_\text{F}}$, $\alpha=\arccos(E/\Delta)$ for $E<\Delta$. Here, the wavefunction is obtained under the condition of $\mu_\text{SC}$ is much larger than $\Delta$. $b$ and $c$ are coefficients of wavefunctions that are coherent superpositions of electron and hole excitations, and can be obtained by solving the continuity condition at the interface. In our consideration, chemical potentials and magnetizations in ferromagnetic insulators are respectively set to be zero and $\Delta$.

Spin-pairing correlations can be obtained from the retarded Green's function~\cite{X.Liu2015,N.Yuan2015a,L.Gorkov2001,F.Crepin2015}, which is closely related to the spectral function $A(E, x)=\psi(x)\otimes\psi^\dagger(x)$ with $\psi(x)$ being the wave function of the bound state~\cite{X.Liu2015}. And off-diagonal block $A^{\text{off}}(E,x)$ can be expressed as:
\begin{align}
	A^\text{off}(E,x)=
	\begin{pmatrix}
		0 & d_{0}\sigma_{0}+{\boldsymbol{d}}\cdot {\boldsymbol{\sigma}}\\
		d_{0}^{\ast}\sigma_{0}+\boldsymbol{d}^{\ast}\cdot \boldsymbol{\sigma} & 0
	\end{pmatrix},
\end{align}
where $\sigma_{0}$ is $ 2\times2 $ identity matrix, $ d_{0} $ and $ \boldsymbol{d} $ represent separately amplitudes of the spin-singlet and spin-triplet pairing correlations. To be specific, $ f_{0}=d_{0} $ is the pairing amplitude of spin-singlet correlation $ \ket{\uparrow\downarrow}-\ket{\downarrow\uparrow} $; $ f_{1}=-d_{x}+\mathrm{i}d_{y} $, $ f_{2}=d_{x}+\mathrm{i}d_{y} $ and $ f_{3}=d_{z} $ are respectively the pairing amplitudes of spin-triplet correlations $\ket{\uparrow\uparrow}$, $ \ket{\downarrow\downarrow} $ and $\ket{\uparrow\downarrow}+\ket{\downarrow\uparrow} $.

\begin{figure}
	\centering
	\includegraphics[width=8cm]{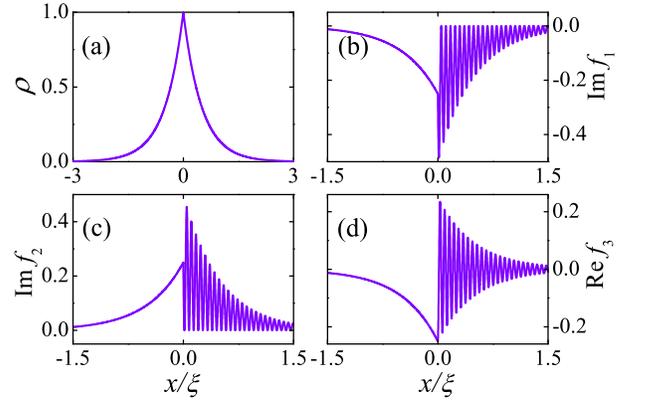}
	\caption{(color online) (a) Probability density $\rho$ of the zero-energy MBS as a function of $x$, with $x=0$ the interface of FI-SC junction. (b-d) Plot of spin-triplet pairing correlations $\ket{\uparrow\uparrow}, \ket{\downarrow\downarrow}$ and $\ket{\uparrow\downarrow}+\ket{\downarrow\uparrow}$ in FI-SC junction, respectively. Here, chemical potential is $\mu_\text{SC}=50\Delta$, and superconducting coherence length is defined as $\xi=\hbar\upsilon_\text{F}/\Delta$.}\label{fig2}
\end{figure}

\emph{Single Majorana Bound State---.} In a FI-SC junction as displayed in Fig.~\ref{fig1}(a), a zero-energy MBS can be formed on the boundary of two-dimensional topological insulator where the gapless edge modes can be drove to open up a band gap by either ferromagnetism or $s$-wave superconducting pair potential due to its spin-momentum locking. The probability density $\rho(x)=\psi^\dagger(x)\psi(x)$ of this bound state is plotted in Fig.~\ref{fig2}(a) as a function of position $x$ where one can find that $\rho$ decays exponentially with the increase of $|x|$ indicating that the MBS is localized around the interface. We further calculate the spectral function and find a vanishing value of $d_0$. Thus, there is no spin-singlet pairing correlation. However, spin-triplet pairing correlations present since $\bf{d}$ is not zero whose component $d_{x}$ is a pure imaginary number while $d_{y,z}$ are real numbers. Therefore, the spin-triplet pairing correlation amplitudes $f_{1,2}= \mp d_{x}+\mathrm{i}d_{y}$ have only imaginary parts while $f_{3}=d_z$ has only real part as plotted in Figs.~\ref{fig2}(b)-\ref{fig2}(d) where these amplitudes are also localized around the interface and exhibit the Friedel-type spatial oscillation in the $x>0$ superconducting region with a periodicity of $1/k_\text{SC}$.

\begin{figure}
	\centering
	\includegraphics[width=8cm]{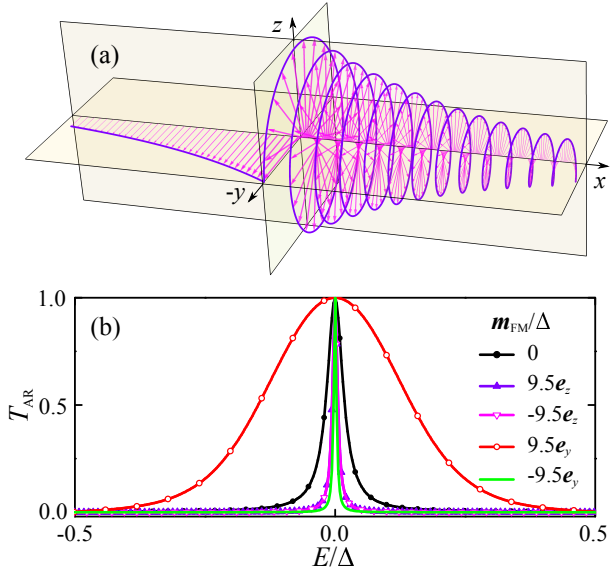}
	\caption{(color online) (a) Local spin polarization of the MBS $ \boldsymbol{s}(x)$ changes as a function of position $x\in[-0.7\xi,0.8\xi]$, which is in the $y-z$ plane. (b) Probability of Andreev reflection $ T_\text{AR} $ as a function of excitation energy $ E $ of the incident electron for different magnetization $ \boldsymbol{m}_{\text{FM}} $. Here, we choose $ l_\text{FI}=2\xi $, $ \mu_\text{FM}=10\Delta $ and $ \mu_\text{SC}=50\Delta $.}\label{fig3}
\end{figure}

The existence spin-triplet pairing correlation indicates that the MBS has non-zero spin polarization $s_{i}(x)=\psi^\dagger(x) (\tau_{z}\otimes\sigma_{i}) \psi(x)$ where $i=x,y,z$ and $ \tau_{z} $ describes the particle-hole degree of freedom~\cite{H.Ebisu2015, D.Sticlet2012}. We find that the $x$-component of spin polarization $ \boldsymbol{s}(x) $ vanishes in the whole regime while $ z $-component is also zero in the ferromagnetic insulator region as shown in Fig.~\ref{fig3}(a) where the $y$ and $z$ components are plotted as a function of $ x $. It shows that, in the superconductor region of $x>0$, the local spin polarization of the MBS varies dramatically and exhibits a spin helix structure; while in the ferromagnetic insulator region (i.e. $x<0$) the spin polarization orients along $ -y $ direction, perpendicular to the magnetization $ \boldsymbol{m} $ of ferromagnetic insulator. We notice that, to our surprise, the local spin-polarization direction of MBS is the same as that of the Cooper pair near the interface, which can be obtained by $ \boldsymbol{s}_{\rm{C}}(x) = \mathrm{i} (\boldsymbol{d} \times \boldsymbol{d}^{\ast})/|\boldsymbol{d}|^{2} $ where subscript $C$ is employed to distinguish this quantity from the spin polarization of MBS $\boldsymbol{s}(x)$~\cite{A.Leggett1976}. This feature strongly suggests that MBS and spin-triplet Cooper-pair correlations across the interface are two aspects of the same thing.

After understanding the spin-pairing correlations and spin polarization of the MBS, one can naturally determine the spin-related transport property. For example, the spin-triplet pairing correlations can result in the selective equal-spin Andreev reflection, and the orientation of spin polarization of MBS further dominates the selective direction. Therefore, spin-related properties of MBS in FI-SC can be experimentally detected by using a transport setup displayed in Fig.~\ref{fig1}(c), where a ferromagnetic metal (FM) lead is connected to the FI-SC junction.  The ferromagnetic metal is formed on the edge of two-dimensional topological insulator where the ferromagnetism can be induced by the proximity effect of a ferromagnetic insulator and the metallic state can be induced by tuning the Fermi energy as indicated in Fig.~\ref{fig1}(d). Moreover, the magnetization of the ferromagnetic metal is controllable and can be changed to any direction $ \boldsymbol{m}_{\text{FM}}=(m_{x},m_{y},m_{z}) $. In order to utilize the electronic transport property in the junction, the length of ferromagnetic insulator should be carefully chosen. In our consideration, we set $l_{\rm{FI}}=2\xi$. Figure~\ref{fig3}(b) displays the probability of Andreev reflection $ T_\text{AR} $ as a function of excitation energy $ E $ for different $ \boldsymbol{m}_{\text{FM}} $. One can see that the zero-energy MBS leads to resonant Andreev reflection at $ E=0 $, with a finite width of peak due to the weak coupling between the lead and superconductor through the ferromagnetic insulator (cf. Fig.~\ref{fig1}(d)).

It is noteworthy that the resonant Andreev reflection occurs for different $ \boldsymbol{m}_{\text{FM}} $ since it is determined by a joint effect between the spin-flip scattering and equal spin Andreev reflection occurring in ferromagnetic insulator regime as described below. In the lead, the spin of  incident electron parallels to $ (\hbar\upsilon_\text{F}k^{+}+m_{x},m_{y},m_{z})^{T} $, while that of Andreev reflected hole parallels to $ (-\hbar\upsilon_\text{F}k^{-}+m_{x},m_{y},m_{z})^{T} $, where $ k^{\pm}=\left[\sqrt{(\mu_\text{FM}\pm E)^2-m_{z}^2-m_{y}^2}\mp m_{x}\right]/\hbar\upsilon_\text{F} $. For vanishing $ \boldsymbol{m}_{\text{FM}}$, spins of zero-energy incident electron and reflected hole in the lead are respectively along $+x $ and $-x$ axis, but in the ferromagnetic insulator, these spins are flipped to $ y $ axis. Due to the spin-flip scattering, the resonant Andreev reflection exhibits a narrow peak as displayed by solid-circled line in Fig.~\ref{fig3}(b). For a very large $ \boldsymbol{m}_{\text{FM}} $ pointing along $+y$ axis, spins of zero-energy incident electron and reflected hole in the lead are both approximately along $ +y $ axis, equal spin Andreev reflection occurs for the weakest spin-flip scattering in the absence of spin flip, and the resonant Andreev reflection exhibits the widest peak as displayed by empty-circled line in Fig.~\ref{fig3}(b). As a comparison, for large $ \boldsymbol{m}_{\text{FM}} $ pointing along $ -y $ axis, the strongest spin-flip scattering occurs in the ferromagnetic insulator, leading to the narrowest peak of resonant Andreev reflection as displayed by green line in Fig.~\ref{fig3}(b). These remarkable transport signatures can be utilized to verify the presence of spin-triplet correlations and determine the spin polarization of MBS.

The spin-selective Andreev reflection in our system is different from that induced by the MBS in the vortex core of a topological superconductor, semiconductor-superconductor heterostructures or magnetic atomic chains on superconductors, where spin-selective Andreev reflection occurs only when the spin orientation of incident electron is parallel to that of MBS{~\cite{H.H.Sun2016, L.H.Hu2016}. Because in FI-SC junction the spin polarization of MBS in ferromagnetic insulator region is determined by the evanescent wavefunction in $-x$ direction as given in above.
While in FM-FI-SC junction, the wavefunction of incident electron in ferromagnetic metal lead moving along $x$ direction is matched with the an evanescent wavefunction moving in $x$ direction in ferromagnetic insulator, which is $\psi^{\prime}_\text{FI}(x)=n_\text{e}(\hbar\upsilon_\text{F}k^{+}_\text{FI},E+\mu_\text{FI}-m_\text{L},0,0)^{T}\mathrm{e}^{\mathrm{i}k^{+}_\text{FI}x}
+n_\text{h}(0,0,-\hbar\upsilon_\text{F}k^{-}_\text{FI},E-\mu_\text{FI}-m_\text{L})^{T}\mathrm{e}^{\mathrm{i} k^{-}_\text{FI}x}$.
Such evanescent wavefunction has opposite spin polarization compared to that of MBS wavefunction, also in fact the MBS wavefunction is the reflected wavefunction of $\psi^{\prime}_\text{FI}(x)$ in FM-FI-SC junction.

\begin{figure}
	\centering
	\includegraphics[width=8cm]{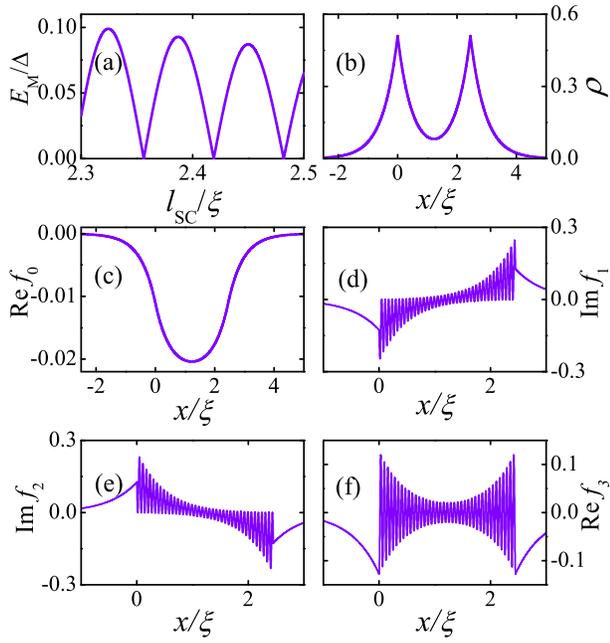}
	\caption{(color online) (a) Coupling energy $ E_\text{M} $ of two MBSs as a function $ l_\text{SC} $. (b) Probability density of the fermionic state formed by two coupled MBSs as a function of $ x $, with $ x=0, 2.45\xi $ being the two interfaces of FI-SC-FI junction. (c-f) Amplitudes of spin-pairing correlations $ \ket{\uparrow\downarrow}-\ket{\downarrow\uparrow}, \ket{\uparrow\uparrow}, \ket{\downarrow\downarrow} $ and $ \ket{\uparrow\downarrow}+\ket{\downarrow\uparrow} $, respectively. Here, $ \mu_\text{SC}=50\Delta $.}\label{fig4}
\end{figure}

\emph{Two Coupled Majorana Bound States---.} Now, we move to the system with coupling between two MBSs at two interfaces of superconductor in FI-SC-FI junction as displayed in Fig.~\ref{fig1}(b). When the length of superconductor $l_\text{S}$ is finite, the wavefunctions of two MBSs are overlapped and coupled to each other with a coupling energy of $E_\text{M} $, which split the two zero-energy MBSs into two fermionic states of energies $ \pm E_\text{M} $. As displayed in Fig.~\ref{fig4}(a), $E_\text{M} $ decreases and oscillates with the increase of the length of superconductor $ l_\text{SC}$~\cite{J.Nilsson2008, S.Sarma2012}. Such oscillation can be approximated as $ E_\text{M} \varpropto \mathrm{e}^{-l_\text{SC}/\xi} \cos(\mu_\text{SC}l_\text{SC}/\hbar\upsilon_\text{F}) $, which implies that the coupling energy also oscillates as the increase of $\mu_\text{SC}$. We further plot the probability density $ \rho(x) $ for the fermionic state of $E_\text{M}$ in Fig.~\ref{fig4}(b) as a function of $x$, where one can clearly see that $\rho$ reaches the maxima at two interfaces indicating the non-locality of the wavefunction. This character also manifest itself for the fermionic state of $-E_\text{M}$.

Different from the single MBS, these two fermionic states possess a weak spin-singlet pairing correlation as shown in Fig.~\ref{fig4}(c), which takes the maximum at the center of superconductor and is much smaller than spin-triplet ones shown in Figs.~\ref{fig4}(d)-\ref{fig4}(f) where $ f_{1,2} $ are also pure imaginary while $ f_{3} $ is real. From these figures, one can find that the spin-triplet pairing correlations are maximized around both $ x=0 $ and $ x=l_\text{SC} $ and oscillate spatially in the superconductor, partly showing the characteristics of two decoupled MBSs. We further explore the local spin polarizations $ \boldsymbol{s}(x) $ of the fermionic states, and find that they also have only finite $ y $-component $ s_{y} $ in the two ferromagnetic insulator regions and has no $ x $-component $ s_{x} $ in the superconductor region. Figure~\ref{fig5}(a) displays $ s_{y}(x) $ and $ s_{z}(x) $ as a function of $ x $. It shows that the local spin polarization direction changes spatially manifesting itself as a spin helix in the superconductor region, while is fixed along $ \mp y $-direction in the left and right ferromagnetic insulator and meanwhile perpendicular to the magnetization $ \boldsymbol{m} $ in both ferromagnetic insulators.

In contrast to the single MBS, the Andreev reflection probability $ T_\text{AR} $ shows strong tunability by changing the chemical potential of superconductor $ \mu_\text{SC} $. In Fig.~\ref{fig5}(b), we plot $ T_\text{AR} $ as function of $ \mu_\text{SC} $ and energy $ E $ of the incident electron, where we find that, given $\mu_{\rm{SC}}$, the resonant peak occurs when $E$ is close enough to $\pm E_{\text{M}}$ as indicated by the black dashed line. Therefore, the energy spacing of these two peaks of $ T_\text{AR} $ corresponds to twice of the coupling energy $2E_{\text{M}}$. By changing $\mu_{\rm{SC}}$, one can find that the energy spacing of these two resonance peak oscillates as shown in Fig.~\ref{fig5}(b), which is in agreement with the dependence of $E_{\rm{M}}$ on $\mu_{\rm{SC}}$. This dependence of $ T_\text{AR} $ on $\mu_\text{SC}$ provides an unambiguous evidence for the existence of MBSs by the measurement of charge conductance. Moreover, the width of the resonance peak is also spin-orientation dependent as shown in Fig.~\ref{fig5}(c), which displays the probability of Andreev reflection $T_\text{AR}$ in the FM-FI-SC-FI junction. It is found that when $ \boldsymbol{m}_{\text{FM}} $ is positively large along the $y$ direction, $T_\text{AR}$ provides the widest peak as shown by the red dotted line; while when $ \boldsymbol{m}_{\text{FM}} $ is negatively large along the $-y$ direction, $T_\text{AR}$ gives the narrowest peak as shown by the green curve. The dependence of $T_\text{AR}$ on the magnetization of the ferromagnetic lead originates from the same physics as that for the FM-FI-SC junction. Therefore, the splitting of $T_\text{AR}$ peak and its dependence on the magnetization can be utilized to detect the spin-triplet correlations and spin polarization of coupled MBSs, e.g., by directly measuring the differential conductance in experiment.

\begin{figure}
	\centering
	\includegraphics[width=8cm]{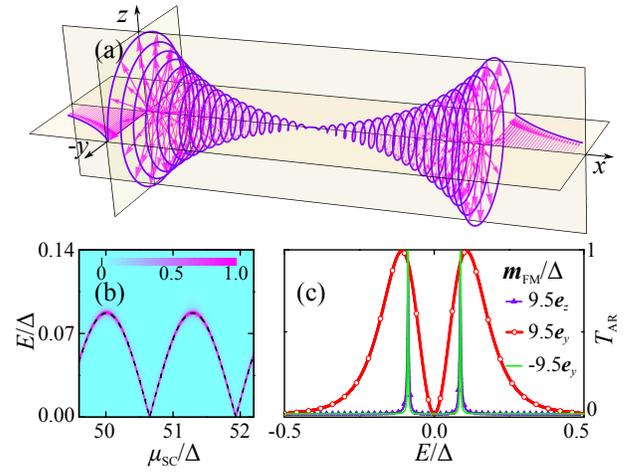}
	\caption{(color online) (a) Local spin polarization of the fermionic state formed by two coupled MBSs $ \boldsymbol{s}(x)$ as a function of position $x\in[-0.5\xi,3\xi]$, with $\mu_\text{SC}=50\Delta$. (b) Probability of Andreev reflection $T_\text{AR}$ as a function of excitation energy $E$ of the incident electron and $\mu_\text{SC}$, with the black-dashed line being the coupling energy of the two MBSs. (c) $T_\text{AR}$ as a function of $E$ for different magnetization of $ \boldsymbol{m}_{_\text{FM}} $ with  $\mu_\text{SC}=50\Delta$. In (b-c), $l_\text{FI}=2\xi$ and $\mu_\text{FM}=10\Delta$.}\label{fig5}
\end{figure}

\emph{Conclusions---.} In summary, we show that a single zero-energy MBS at the boundary of a two-dimensional topological insulator exhibits three types of spin-triplet pairing correlations, while for two coupled MBSs there exists an additional weak spin-singlet pairing correlation. The dominated spin-triplet pairing correlations lead to nonzero spin-polarization of MBSs, which oscillates in superconducting regime while remains the same in ferromagnetic insulator regime. We find that resonance peak of selective equal-spin Andreev reflection occurs for incident electrons with any spin orientation, which however strongly influences the width of this resonance peak. A widest (narrowest) resonance-peak width occurs when an zero-energy incident electrons with spin-polarization (anti-)parallel to that of MBS in ferromagnetic insulator regime. For the coupled MBSs, the resonance peak at zero energy splits into two and the splitting oscillates periodically as the the chemical potential of superconductor changes. These phenomena can serve as unambiguous signals for experimental verification of the presence of spin-triplet correlations and to determine the spin polarization of MBSs.

\emph{Acknowledgments---.} This work was financially supported by the National Key R \& D Program (2016YFA0301700), NNSFC (11474265), the China Government Youth 1000-Plan Talent Program, Fundamental Research Funds for the Central Universities (WK3510000001 and WK2030020027), and China Postdoctoral Science Foundation (2016M590569). The Supercomputing Center of USTC is gratefully acknowledged for the high-performance computing assistance. We acknowledge Xin Liu and Qiang Cheng for helpful discussions.

\end{document}